\documentclass[12pt]{article}
\usepackage{amsmath,cite,epsfig,a4,times,euler,euscript}
\usepackage{graphics}
\usepackage[usenames]{color}
\renewcommand{\baselinestretch}{1.1}
\newcommand{\myTitle}[1]{\begin{center}{\bf\Huge #1}\\[5ex]\end{center}}
\newcommand{\myAuthor}[1]{\begin{center}{\Large #1}\\[2ex]\end{center}}
\newcommand{\myAffiliation}[1]{\\[1ex]{\it\large #1}}
\newcommand{\myEmail}[1]{}
\newcommand{\myDate}{\begin{center}{\large\today}\\[5ex]\end{center}}
\newcommand{\myAbstract}[1]{\begin{center}\renewcommand{\baselinestretch}{1}{\bf Abstract}\\[2ex]\parbox{0.8\linewidth}{\small\hspace{15pt} #1}\end{center}\vspace{\baselineskip}}
\newcommand{\myReport}[1]{\hspace{\fill} #1}
\newcommand{\myPreprint}[1]{}
\newcommand{\myKeywords}[1]{}
\newcommand{\myFigure}[1]{\begin{figure}[t]\begin{center}#1\end{center}\end{figure}}

\newcommand{\lid}[2]{#1\!\cdot\!#2}
\newcommand{\slashp}{p\hspace{-6.5pt}/}

\newcommand{\slashl}{\ell\hspace{-6.0pt}/}
\newcommand{\slashk}{k\hspace{-6.5pt}/}
\newcommand{\slashe}{\varepsilon\hspace{-6.0pt}/}
\newcommand{\Appendix}[1]{Appendix~\ref{#1}}

\newcommand{\Figure}[1]{Fig.~\ref{#1}}
\newcommand{\Equation}[1]{Eq.~(\ref{#1})}

\newcommand{\ie}{{\it i.e.}}

\newcommand{\eg}{{\it e.g.}}
\newcommand{\bram}[1]{\langle#1|}
\newcommand{\brap}[1]{[#1|}
\newcommand{\ketm}[1]{|#1]}
\newcommand{\ketp}[1]{|#1\rangle}

\newcommand{\bkmp}[2]{\langle#1|#2\rangle}
\newcommand{\bkpm}[2]{[#1|#2]}

\newcommand{\UU}[2]{\langle#1#2\rangle}
\newcommand{\VV}[2]{[#1#2]}
\newcommand{\srac}[2]{{\textstyle\frac{#1}{#2}}}

\newcommand{\imag}{\mathrm{i}}
\newcommand{\xone}{x_1}
\newcommand{\xtwo}{x_2}

\newcommand{\vep}{\varepsilon}
\newcommand{\gQCD}{g_\mathrm{S}}
\newcommand{\uq}{u}
\newcommand{\ub}{\bar{u}}

\begin{document}

\myReport{IFJPAN-IV-2013-15}
\myPreprint{}

\myTitle{%
Scattering amplitudes\\[0.5ex] with off-shell quarks%
}

\myAuthor{%
A.~van~Hameren$^\textit{\,a}$, K.~Kutak$^\textit{\,a}$ and T.~Salwa$^\textit{\,a,b}$%
\myAffiliation{%
$^\textit{a}$%
The H.\ Niewodnicza\'nski Institute of Nuclear Physics\\
Polish Academy of Sciences\\
Radzikowskiego 152, 31-342 Cracow, Poland\\[1ex]
$^\textit{b}$%
Faculty of Physics, Astronomy and Applied Computer Science\\
Jagiellonian University\\[0.5ex]
Reymonta 4, 30-059 Cracow, Poland%
\myEmail{hameren@ifj.edu.pl}
}
}

\myDate

\myAbstract{%
We present a prescription to calculate manifestly gauge invariant tree-level scattering amplitudes for arbitrary scattering processes with off-shell initial-state quarks within the kinematics of high-energy scattering.
}

\myKeywords{QCD}

%

\section{Introduction\label{Sec:intro}}
Factorization procedures are the key to the application of perturbative QCD to the calculation of cross sections of hard scattering processes at the Large Hadron collider.
They follow the physical picture of the parton model, in which it is eventually the quarks and gluons, or {\em partons\/}, inside the colliding hadrons that interact in the hard scattering process.
In {\em collinear factorization\/}, these partons simply transport a fraction of the momentum of the hadron into the scattering process, encoded by a single proportionality factor usually denoted by $x$.
The factorization happens via the convolution of a process-dependent partonic cross section with universal parton density functions (pdfs), which only depend on this $x$ variable and an unphysical scale, the factorization scale.
The total cross section is independent of this scale if all orders in perturbation theory are taken into account, and in~\cite{Ellis:1978ty} it has been proven that this indeed works and that the procedure is consistent.

One may then wonder whether it is possible to generalize the factorization procedure, and allow for the partons to carry momentum components into the partonic scattering process that are independent of the original hadron momentum, or more specifically, that are transverse.
This would allow for calculations at leading order in QCD to incorporate kinematical effects that only appear at higher orders within collinear factorization.

The extension of collinear factorization to allow for inclusion transverse momentum effects goes under a name {\em Transversal Momentum Dependent} (TMD) factorization (we refer the reader to \cite{Mulders:2011zt} and references therein) and is in principle valid at large $x$.
It has been realized in \cite{Gribov:1984tu,Catani:1990eg} that the introduction of the extra momentum components implies an extra energy scale which may be assumed to be much lower than the total energy of hadronic scattering process.
This approach goes under the name of {\em high-energy factorization}.
It becomes relevant when the transverse components are sizable compared to the longitudinal components carried into the partonic process, \ie\ for low values of $x$.

The partonic cross section within high-energy factorization requires matrix elements with off-shell initial-state partons and needs to be convoluted with parton densities which depend on longitudinal and transverse momentum \cite{Kuraev:1976ge,Kuraev:1977fs,Balitsky:1978ic,Ciafaloni:1987ur,Catani:1989sg,Catani:1989yc,Balitsky:1995ub, Kovchegov:1999yj,Kovchegov:1999ua,Kutak:2011fu,Kutak:2012yr,Kutak:2012qk}.
Various approaches to calculate the matrix elements for off-shell initial-state gluons exist~\cite{Lipatov:1995pn,Antonov:2004hh,vanHameren:2012uj,vanHameren:2012if}.
The main issue is to ensure that they are gauge invariant and that they satisfy the necessary Ward identities.
While far less popular in literature, approaches for off-shell quarks have also been developed.
In particular, the effective action approach for off-shell gluons~\cite{Lipatov:1995pn} has also been applied for off-shell quarks~\cite{Lipatov:2000se}, and this approach has been followed \eg\ in~\cite{Bogdan:2006af}.
Recently, it has been used to calculate scattering amplitudes for several scattering processes in~\cite{Nefedov:2013ywa}.
Other recent work involving off-shell quarks can be found in~\cite{Hautmann:2012sh,Hautmann:2012pf,Saleev:2008rn,Kniehl:2008qb,Ermolaev:2011aa}.

In this paper, we will derive a prescription to calculate manifestly gauge invariant tree-level scattering amplitudes with off-shell initial-state quarks and an arbitrary number of final-state particles.
It is essentially the generalization of the work in~\cite{vanHameren:2012if} for quarks, and can straightforwardly be implemented in numerical programs using the well-known efficient methods for tree-level calculations.
Whereas prescription in~\cite{vanHameren:2012if} for gluons was shown to be equivalent to the effective action approach of~\cite{Lipatov:1995pn}, the prescription for quarks presented here will be shown to be equivalent to the effective action approach of~\cite{Lipatov:2000se}.

\section{Construction}
In~\cite{vanHameren:2012if}, a pair of auxiliary quarks was introduced enabling the embedding of the process $g^*g^*\to X$ with the off-shell gluons into the process $q_Aq_B\to q_Aq_B\,X$ with on-shell quarks, and where $X$ represents any set of final-state particles.
It was shown that by applying eikonal Feynman rules to the quark lines, gauge invariant scattering amplitudes are obtained that are defined in the desired kinematical configuration that is relevant in high-energy factorization.
This configuration is such, that the off-shell initial-state partons carry momenta
%
\begin{equation}
k_1^\mu = \xone\ell_1^\mu + k_{1\perp}^\mu
\quad,\quad
k_2^\mu = \xtwo\ell_2^\mu + k_{2\perp}^\mu
~,
\end{equation}
%
where $\ell_{1},\ell_2$ are the light-like momenta associated with the colliding hadrons, and where $k_{1\perp},k_{2\perp}$ are transverse to both $\ell_1$ and $\ell_2$.
The momentum fractions $\xone,\xtwo$ are between $0$ and $1$.

It was stressed in~\cite{vanHameren:2012if} that the formalism works for two off-shell gluons, but it was also clear that the treatment of the gluons is completely independent, and that the formalism can be viewed as a trivial generalization from one off-shell gluon to two off-shell gluons.
Here, we will consider the case of a single off-shell initial-state quark, and it will be clear that the situation can be trivially generalized to the case of two off-shell initial-state partons, be it quarks and/or gluons.
So we consider the process
%
\begin{equation}
u^*g\to u\,X
~,
\end{equation}
%
where $X$ represent an arbitrary, but definite, set of final-state particles.
\myFigure{
\epsfig{figure=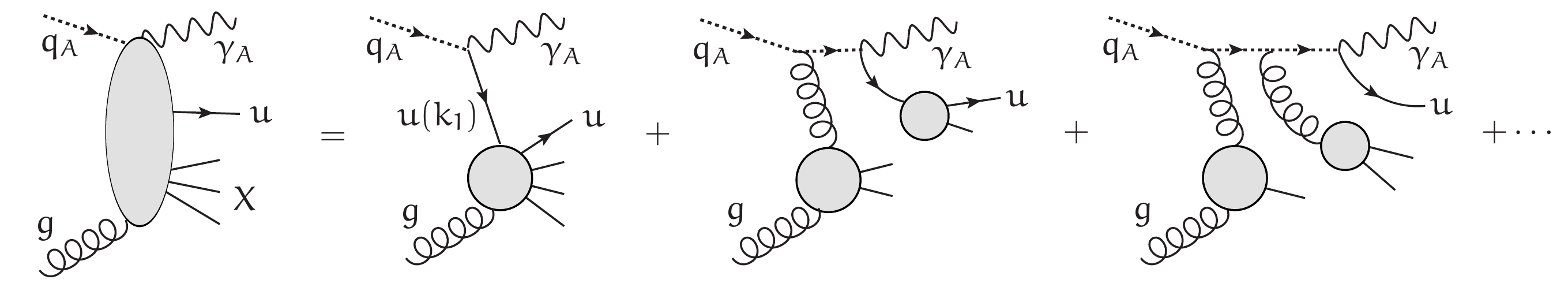,width=0.99\linewidth}
\caption{\small\label{Fig1}A few terms in the classification of the graphs contributing to $q_A g\to \gamma_A u\,X$ w.r.t.\ the gluons attached to the quark line.}
}
We will follow the strategy of embedding this process into a larger on-shell proces again.
To this end, we introduce an auxiliary quark $q_A$ and an auxiliary photon $\gamma_A$ which interact with the $u$-quark via the vertex
%
\begin{equation}
\raisebox{-16pt}{\epsfig{figure=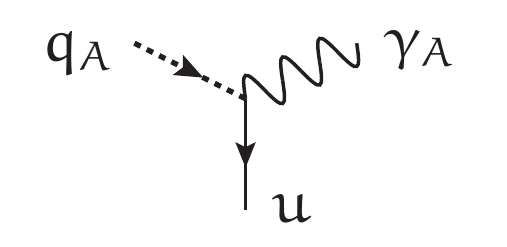,width=90pt}} =\; -\imag\gamma^\mu
~.
\label{qAuvertex}
\end{equation}
%
The auxiliary photon does not interact with any other particles, and can be interpreted as allowing for neutral flavor-changing currents involving $q_A$ and $u$ quarks.
It is a color singlet, whereas the quark is in the fundamental representation.
The quark further only interacts with gluons via normal quark-gluon vertices.
Now, we consider the process
%
\begin{equation}
q_A g \to \gamma_A u\,X
~.
\end{equation}
%
\Figure{Fig1} shows a classification of graphs contributing to the process.
The first term on the r.h.s.\ contains all the graphs with the off-shell quark with momentum $k_1$.
To arrive at a gauge invariant amplitude, also the other terms have to be taken into account.
The momenta of the auxiliary quark $q_A$ and photon $\gamma_A$ are $p_A$ and $p_{A'}$ respectively, and following the approach of~\cite{vanHameren:2012if}, we assign the values
%
\begin{equation}
p_A^\mu = (\Lambda+\xone)\ell_1^\mu
- \frac{\lid{k_{1\perp}}{\ell_4}}{\lid{\ell_1}{\ell_2}}\,\ell_3^\mu
\quad,\quad
p_{A'}^\mu = \Lambda\ell_1^\mu
+ \frac{\lid{k_{1\perp}}{\ell_3}}{\lid{\ell_1}{\ell_2}}\,\ell_4^\mu
~,
\label{pApAp}
\end{equation}
%
where $\ell_3,\ell_4$ span the transverse space, and are defined by
%
\begin{equation}
\ell_3^\mu=\srac{1}{2}\bram{\ell_2}\,\gamma^\mu\,\ketm{\ell_1}
\quad,\quad
\ell_4^\mu=\srac{1}{2}\bram{\ell_1}\,\gamma^\mu\,\ketm{\ell_2}
~.
\label{eqn:l3l4}
\end{equation}
%
The momenta in \Equation{pApAp} are on-shell and satisfy
%
\begin{equation}
p_A^\mu - p_{A'}^\mu = k_1^\mu = \xone\ell_1^\mu + k_{1\perp}^\mu
~,
\end{equation}
%
as required, for any value of the dimensionless parameter $\Lambda$.
It was argued in~\cite{vanHameren:2012if} that one can directly assign the spinor $\ketm{\ell_1}$ to the auxiliary quark without spoiling gauge invariance:
%
\begin{equation}
q_A \;\leftarrow\; \ketm{\ell_1}
~.
\end{equation}
%
We will do the same here, but with assumption that the helicity of the final-state $u$-quark is negative, so for
%
\begin{equation}
\textrm{$u$-quark}\;\leftarrow\;\bram{p_u}
~,
\end{equation}
%
where $p_u$ is the momentum of the $u$-quark.
We will come back to the case of positive helicity for the $u$-quark later.
For the polarization vector of $\gamma_A$ we use the proper normalization of the vector $\ell_4$:
%
\begin{equation}
\vep_{A'}^\mu = \frac{\sqrt{2}}{\bkpm{\ell_1}{\ell_2}}\,\ell_4^\mu
~.
\label{polvec}
\end{equation}
%
It is a polarization vector the for a photon with momentum $\ell_1$ and auxiliary vector $\ell_2$, and is also valid for momentum $p_{A'}$ since $\lid{p_{A'}}{\vep_{A'}}=0$.

On-shellness of all particles involved in the constructed amplitude ensures its gauge invariance.
It, however, depends on unphysical imaginary momentum components, and the physical amplitude is extracted by taking the limit of $\Lambda\to\infty$.
This limit only affects the $q_A$-propagators.
For a $q_A$-line with momentum $p$ we have
%
\begin{equation}
\frac{\imag\,\slashp}{p^2}
\;\overset{\Lambda\to\infty}{\longrightarrow}\;
\frac{\imag\,\slashl_1}{2\,\lid{\ell_1}{p}}
~.
\end{equation}
%

The obtained amplitude still needs to be matched to the collinear limit when $k_1^2\to0$.
It has to be multiplied by a kinematical factor, which we will now derive.
Firstly, the first set of graphs on the r.h.s.\ of \Figure{Fig1} dominate, since only they contain $k_1^2$ in the denominator.
Let us abbreviate this set by
%
\begin{equation}
\mathcal{A}
=
\brap{\Psi}\,\frac{\imag\,\slashk_1}{k_1^2}\,(-\imag\slashe_{A'})\,\ketm{\ell_1}
=
\frac{1}{k_1^2}\,\brap{\Psi}\,\slashk_1\,\slashe_{A'}\,\ketm{\ell_1}
~,
\end{equation}
%
where $\brap{\Psi}$ represents the blob, excluding the propagator with momentum $k_1$.
Inserting \Equation{polvec}, we get
%
\begin{equation}
\mathcal{A}
=
\frac{1}{k_1^2}\,\brap{\Psi}\,\slashk_1\,\ketp{\ell_1}\bkpm{\ell_2}{\ell_1}\,\frac{\sqrt{2}}{\bkpm{\ell_1}{\ell_2}}
=
-\frac{\sqrt{2}}{k_1^2}\,\brap{\Psi}\,\slashk_1\,\ketp{\ell_1}
~,
\end{equation}
%
and inserting the expansion of $k_1$ in terms of $\ell_1,\ell_3,\ell_4$, we find%
\begin{equation}
\mathcal{A}
=
\frac{\sqrt{2}}{k_1^2}\,\bkpm{\Psi}{\ell_1}\bkmp{\ell_2}{\ell_1}
\,\frac{\lid{k_{1\perp}}{\ell_4}}{\lid{\ell_1}{\ell_2}}
~.
\end{equation}
%
Here, $\bkpm{\Psi}{\ell_1}$ is the amplitude in the collinear limit, modulo a factor $\sqrt{\xone}$.
In the collinear case, the on-shell initial-state quark would get a spinor $\ketm{\xone\ell_1}=\sqrt{\xone}\,\ketm{\ell_1}$.
Now we note that, for real $\ell_1,\ell_2$, the vectors $\ell_3$ and $\ell_4$ are each others complex conjugate, so that
%
\begin{equation}
k_1^2 = k_{1\perp}^2 = -2\,\frac{|\lid{k_{1\perp}}{\ell_4}|^2}{\lid{\ell_1}{\ell_2}}
\label{eqn:kperpsquare}
\end{equation}
%
and, using also $|\bkmp{\ell_2}{\ell_1}|^2=2\lid{\ell_1}{\ell_2}$, we find
%
\begin{equation}
|\mathcal{A}|^2
=
-2\,\frac{\big|\bkpm{\Psi}{\ell_1}\big|^2}{k_1^2}
=
\frac{\big|\bkpm{\Psi}{x_1\ell_1}\big|^2}{-x_1k_1^2/2}
~.
\end{equation}
%
So we conclude that in order to arrive at an amplitude leading to the correct collinear limit, it has to be multiplied with $\sqrt{-x_1k_1^2/2}$.

The derivation above was for a negative-helicity final-state $u$-quark.
For positive helicity, we need to switch the roles of $\ell_3$ and $\ell_4$, and assign $q_A\leftarrow\ketp{\ell_1}$.
The auxiliary photon gets polarization vector $\sqrt{2}\,\ell_3^\mu/\bkmp{\ell_2}{\ell_1}$.
For processes with an off-shell initial-state anti-quark, the derivation works quite analogously, with spinors $\bram{\ell_1},\brap{\ell_1}$ for the auxiliary quark.
The kinematical factor the amplitude has to be multiplied with is the same for all cases.

We summarize the prescription to calculate gauge-invariant tree-level scattering amplitudes with an off-shell initial-state, say $u$-, quark that gives the correct collinear limit:
\paragraph{Prescription}
\begin{enumerate}
\item Consider the embedding of the process, in which the off-shell $u$-quark is replaced by an auxiliary quark $q_A$, and an auxiliary photon $\gamma_A$ is added in final state.
\item The momentum flow is as if $q_A$ carries momentum $k_1$ and the momentum of $\gamma_A$ is identical to $0$.
\item $\gamma_A$ only interacts via \Equation{qAuvertex}, and $q_A$ further only interacts with gluons via normal quark-gluon vertices.
\item $q_A$-line propagators are interpreted as $\imag\slashl_1/(2\lid{\ell_1}{p})$, and are diagonal in color space.
\item Sum the squared amplitude over helicities of the auxiliary photon. For one helicity, simultaneously assign to the external $q_A$-quark and to $\gamma_A$ the spinor and polarization vector
\begin{equation}
\ketm{\ell_1}
\quad,\quad
\frac{\bram{\ell_1}\gamma^\mu\ketm{\ell_2}}{\sqrt{2}\,\bkpm{\ell_1}{\ell_2}}
~,
\end{equation}
%
and for the other helicity assign
%
\begin{equation}
\ketp{\ell_1}
\quad,\quad
\frac{\bram{\ell_2}\gamma^\mu\ketm{\ell_1}}{\sqrt{2}\,\bkmp{\ell_2}{\ell_1}}
~.
\end{equation}
%
\item Multiply the amplitude with $\sqrt{-x_1k_1^2/2}$.
\end{enumerate}
For the rest, normal Feynman rules apply.

\paragraph{}
Some remarks are at order.
Regarding the momentum flow, we stress, as in~\cite{vanHameren:2012if}, that momentum components proportional to $k_1$ do not contribute in the eikonal propagators, and there is a freedom in the choice of the momenta flowing through $q_A$-lines.

Regarding the sum over helicities, one might argue that only one of them leads to a non-zero result for given helicity of the final-state quark, but there may, for example, be several identical such quarks in the final state with different helicities.

In case of more than one quark in the final state with the same flavor as the off-shell quark, the rules as such admit graphs with $\gamma_A$-propagators.
These must be omitted.
They do not survive the limit $\Lambda\to\infty$ in the derivation, since the $\gamma_A$-propagators are suppressed by $1/\Lambda$.

The rules regarding the $q_A$-line could be elaborated further like in~\cite{vanHameren:2012if}, leading to simplified vertices for gluons attached to this line and reducing the numerator of the eikonal propagators to $1$.
Formulated as above, however, the prescription is more straightforward and closer to familiar Feynman rules.
The equivalent rules for off-shell gluons were presented in~\cite{vanHameren:2013gba}.

For off-shell initial-state anti-quarks, the spinors and polarization vectors for the auxiliary particles become
%
\begin{equation}
\brap{\ell_1}
\;\;,\;\;
\frac{\bram{\ell_1}\gamma^\mu\ketm{\ell_2}}{\sqrt{2}\,\bkpm{\ell_1}{\ell_2}}
\qquad\textrm{and}\qquad
\bram{\ell_1}
\;\;,\;\;
\frac{\bram{\ell_2}\gamma^\mu\ketm{\ell_1}}{\sqrt{2}\,\bkmp{\ell_2}{\ell_1}}
~.
\end{equation}
%
The eikonal propagator for the anti-quark carries an extra minus-sign.

Amplitudes with a second off-shell initial-state quark can be constructed introducing a second auxiliary quark $q_B$ and a second auxiliary photon $\gamma_B$.
This photon does not interact with the $q_A$ quark, and $\gamma_A$ does not interact with $q_B$.
The Feynman rules for these auxiliary particles are the same as before, but with the role of $\ell_1$ and $\ell_2$ interchanged.
Amplitudes with an off-shell initial-state quark and an off-shell initial-state gluon can be constructed using the rules presented in~\cite{vanHameren:2013gba} for the off-shell gluon.

\section{Results}
In \Appendix{appA} we reproduce some non-trivial expressions for scattering amplitudes with off-shell quarks given earlier in~\cite{Nefedov:2013ywa}, thereby indicating the equivalence of our approach to the effective action approach of~\cite{Lipatov:2000se}.
As a first new application, we present the helicity amplitudes for the process $u^*g\to ug$.
\myFigure{
\epsfig{figure=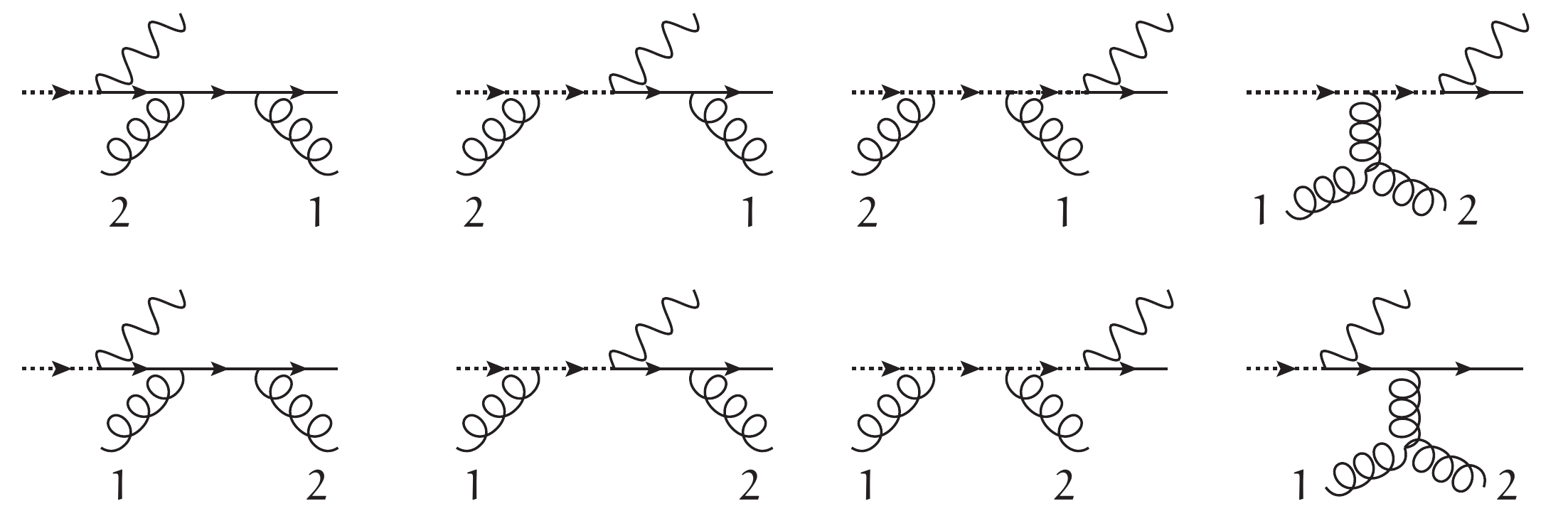,width=0.80\linewidth}
\caption{\small\label{Fig4}Graphs contributing to the process $u^*g\to ug$.}
}
The graphs that have to be taken into account are depicted in \Figure{Fig4}.
%
%
We present the helicity amplitudes with all momenta incoming, \ie\ for the process
\begin{equation}
0 \;\to\; g(p_1)\,g(p_2)\,u(p_\uq)\,\bar{u}^*(p_{\ub}+k_T)
~,
\end{equation}
%
and we denote
%
\begin{equation}
p_{\ub}=-x_1\ell_1
\quad,\quad
k_T=-k_{1\perp}
\quad,\quad
|k_T| = \sqrt{-k_{1\perp}^2}
~.
\end{equation}
%
The amplitude can be decomposed into two color structures following
%
\begin{equation}
\mathcal{M}^{a_1,a_2}_{j_{\uq},j_{\ub}}(1,2,\uq,\ub)
= 2\imag\gQCD^2\Big[
\big(T^{a_1}T^{a_2}\big)_{j_{\uq},j_{\ub}}\mathcal{A}(1,2,\uq,\ub)
+ \big(T^{a_2}T^{a_1}\big)_{j_{\uq},j_{\ub}}\mathcal{A}(2,1,\uq,\ub)\Big]
~,
\end{equation}
where the generators of the color group are normalized such that $\mathrm{Tr}\{T^aT^b\}=\frac{1}{2}\delta^{ab}$.
The non-zero helicity amplitudes ordered with respect to the gluons are then given by
%
\begin{align}
\mathcal{A}(1^+,2^-,\uq^+,\ub^+)&=-
  \frac{\brap{\ub}\slashk_T\ketp{1}}{|k_T|\UU{\ub}{1}}
  \,\frac{\UU{\ub}{1}^3\UU{\uq}{1}}
         {\UU{\uq}{1}\UU{1}{2}\UU{2}{\ub}\UU{\ub}{\uq}}
\quad,
\\
\mathcal{A}(1^-,2^+,\uq^+,\ub^+)&=-
  \frac{\brap{\ub}\slashk_T\ketp{2}}{|k_T|\UU{\ub}{2}}
  \,\frac{\UU{\ub}{2}^3\UU{\uq}{2}}
         {\UU{\uq}{1}\UU{1}{2}\UU{2}{\ub}\UU{\ub}{\uq}}
\quad,
\\
\mathcal{A}(1^+,2^-,\uq^-,\ub^-)&=
  \frac{\bram{\ub}\slashk_T\ketm{1}}{|k_T|\VV{\ub}{1}}
  \,\frac{\VV{\ub}{1}^3\VV{\uq}{1}}
         {\VV{\uq}{1}\VV{1}{2}\VV{2}{\ub}\VV{\ub}{\uq}}
\quad,
\\
\mathcal{A}(1^-,2^+,\uq^-,\ub^-)&=
  \frac{\bram{\ub}\slashk_T\ketm{2}}{|k_T|\VV{\ub}{2}}
  \,\frac{\VV{\ub}{2}^3\VV{\uq}{2}}
         {\VV{\uq}{1}\VV{1}{2}\VV{2}{\ub}\VV{\ub}{\uq}}
\quad,
\\
\mathcal{A}(1^+,2^+,\uq^-,\ub^-)&=-
  |k_T|\,\frac{\UU{\ub}{\uq}^3}
         {\UU{\uq}{1}\UU{1}{2}\UU{2}{\ub}\UU{\ub}{\uq}}
\quad,
\\
\mathcal{A}(1^-,2^-,\uq^+,\ub^+)&=
  |k_T|\,\frac{\VV{\ub}{\uq}^3}
         {\VV{\uq}{1}\VV{1}{2}\VV{2}{\ub}\VV{\ub}{\uq}}
\quad.
\end{align}
In~\cite{vanHameren:2012uj} it was shown that
%
\begin{equation}
\left|\frac{\brap{\ub}\slashk_T\ketp{1}}{|k_T|\UU{\ub}{1}}\right| = 1
~,
\end{equation}
%
so we see that the first four amplitudes are, in terms of the four on-shell momenta, identical to the well-known collinear amplitudes~\cite{Mangano:1990by}, apart from a phase factor.
Notice that also in the color off-diagonal terms in the squared amplitude, \eg\ in $\mathcal{A}(1^+,2^-,\uq^+,\ub^+)\mathcal{A}(2^-,1^+,\uq^+,\ub^+)^*$, the product of the phase factors is $1$, so for the first four helicity configurations, the squared matrix element is given just by the collinear expression evaluated with the on-shell momenta $p_1,p_2,p_u,p_{\ub}$.
This seems counter-intuitive, since these momenta do not satisfy momentum conservation.
Indeed, a matrix element is {\it a priori\/} not unambiguously defined for a set of external momenta that do not satisfy momentum conservation, but a particular explicit expression for that matrix element in terms of the external momenta may be perfectly well defined.

A difference, finally comes with the last two helicity amplitudes, which, contrary to the collinear case, are not identical to zero, but are proportional to $|k_T|$.

\paragraph*{}
The presented prescription to calculate amplitudes with off-shell initial-state quarks has been implemented into a numerical program, and as a second application, we present cross sections for the processes
%
\begin{align}
u\,g &\to d\,\mu^+\,\nu_\mu & u^*\,g &\to d\,\mu^+\,\nu_\mu & u\,g^* &\to d\,\mu^+\,\nu_\mu
\notag\\
u\,g &\to u\,g & u^*\,g &\to u\,g & u\,g^* &\to u\,g 
\notag\\
u\,g &\to u\,g\,g & u^*\,g &\to u\,g\,g & u\,g^* &\to u\,g\,g
\\
u\,g &\to u\,u\,\bar{u}\,g & u^*\,g &\to u\,u\,\bar{u}\,g & u\,g^* &\to u\,u\,\bar{u}\,g
\notag
\end{align}
%
calculated with the help of a toy model unintegrated pdf closely following~\cite{vanHameren:2012if}, namely
%
\begin{equation}
F(x,k_\perp) = f_a(x,\mu)\,\frac{\theta(\mu^2-k_\perp^2)}{Q_0^2\,g(x)}\,\exp\left(-\frac{k_\perp^2}{2Q_0^2\,x\,g(x)}\right)
~.
\label{eqn:defUnint}
\end{equation}
%
The index $a$ indicates the nature of the off-shell parton, and $k_\perp$ now denotes a two-dimensional transverse vector and not its four-dimensional embedding in Minkowski space.
\Equation{eqn:defUnint} is to be understood as a tool to study the off-shell matrix elements, in particular their collinear limits.
$Q_0$ determines the typical scale of the transverse momentum components, and for small values the pdf reduces to the collinear pdf
%
\begin{equation}
\lim_{Q_0\to0}\int\frac{d^2k_{\perp}}{2\pi}\,F(x,k_\perp) = x\,f_a(x,\mu)
~.
\end{equation}
%
The aim of the exercise is to see the difference in behavior between the matrix elements for the off-shell quark and the off-shell gluon.
Therefore, we use the same function $g(x)$ for both types of partons, namely the collinear gluon pdf.

\myFigure{
\epsfig{figure=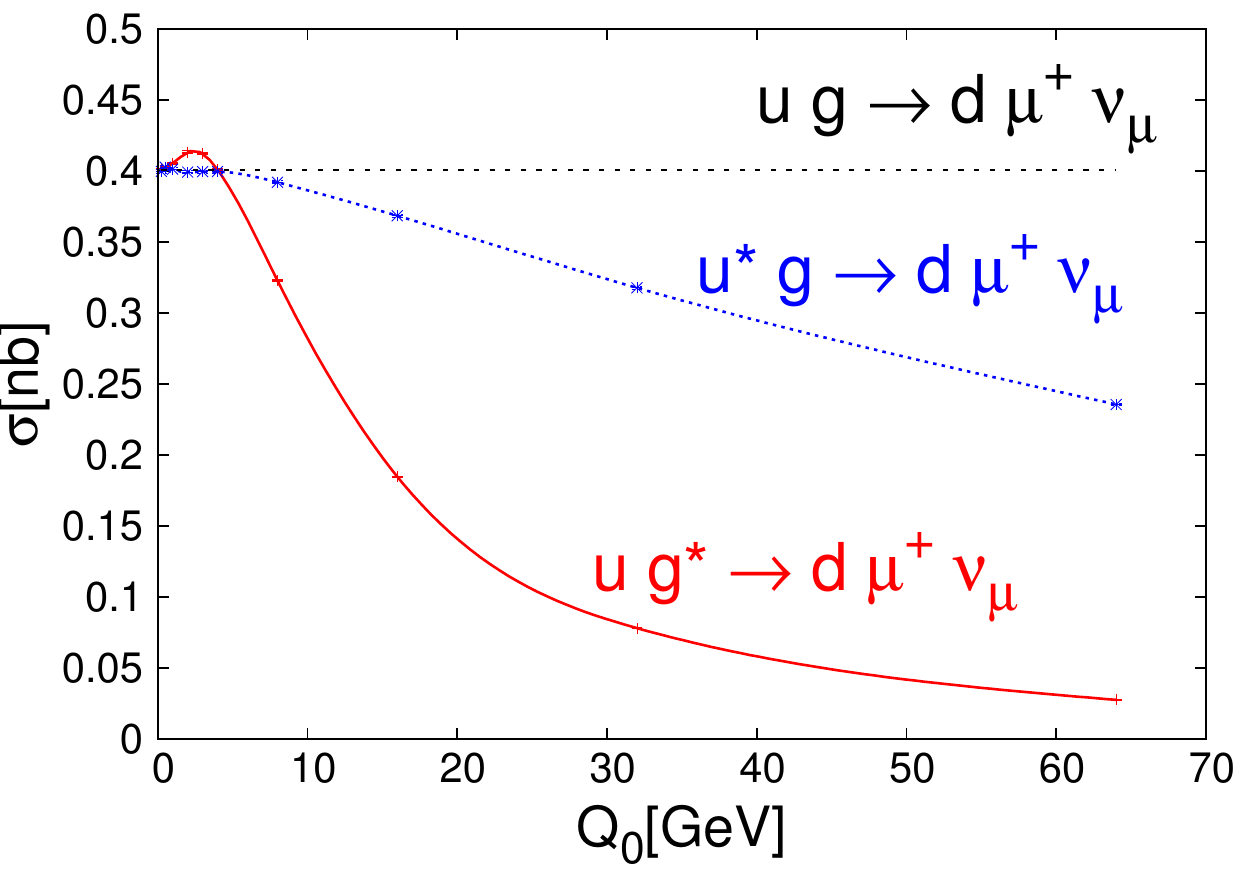,width=0.45\linewidth}
\epsfig{figure=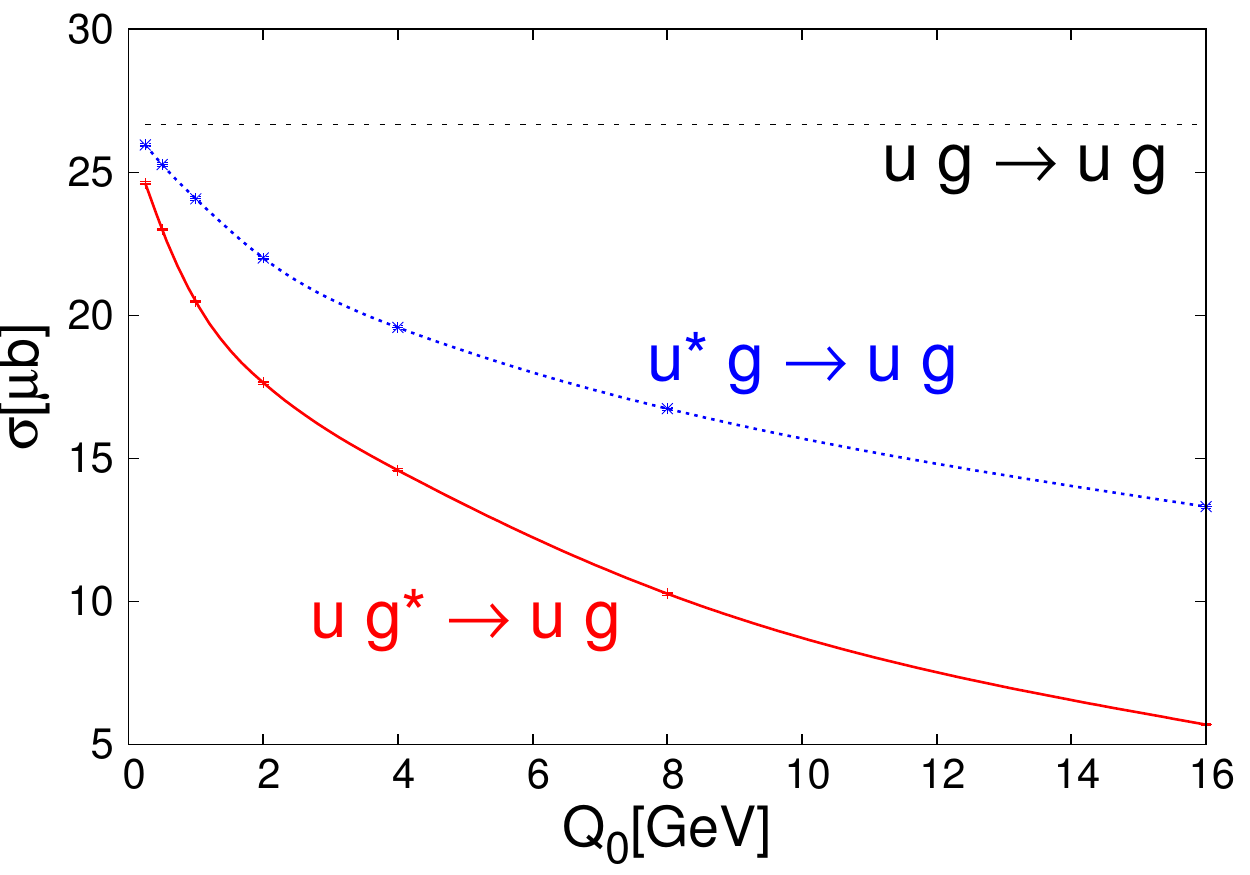,width=0.45\linewidth}\\
\epsfig{figure=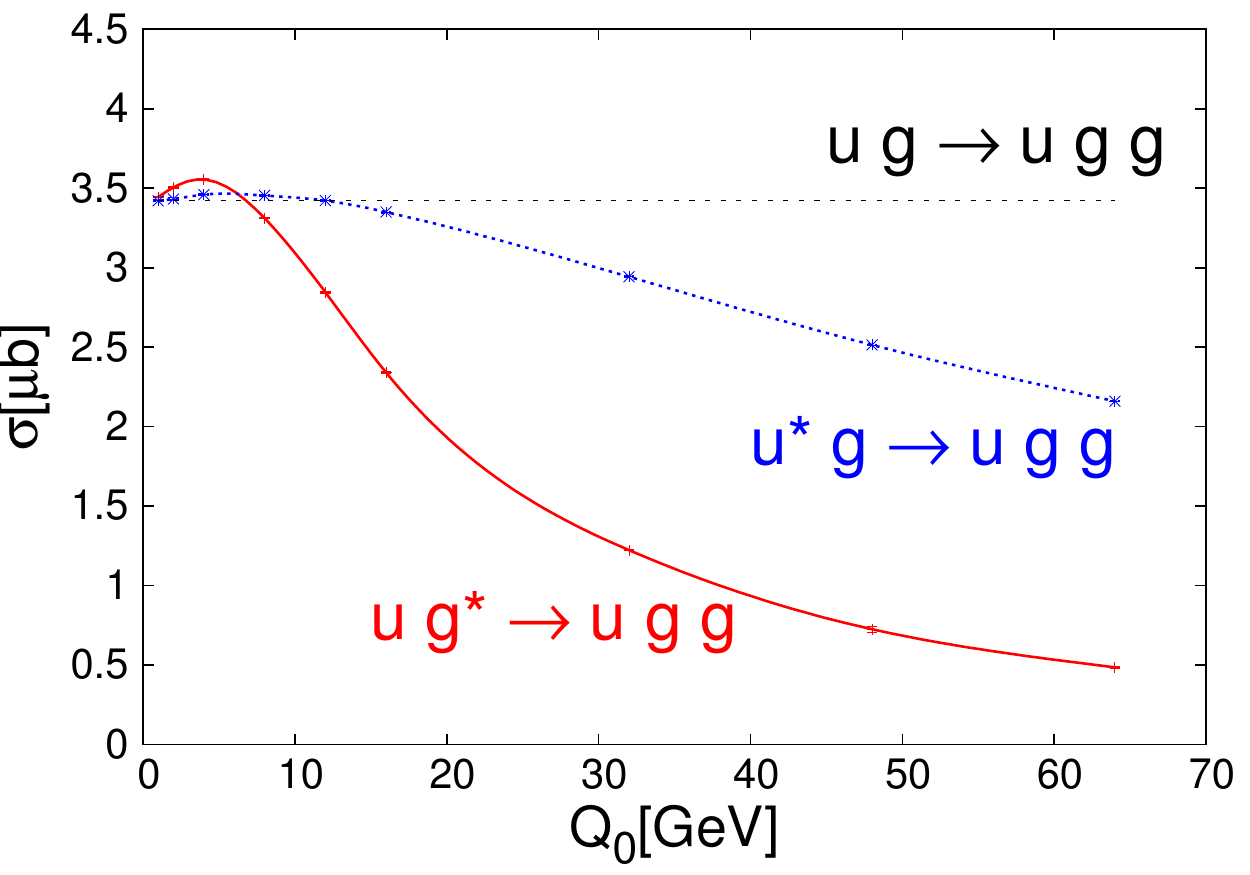,width=0.45\linewidth}
\epsfig{figure=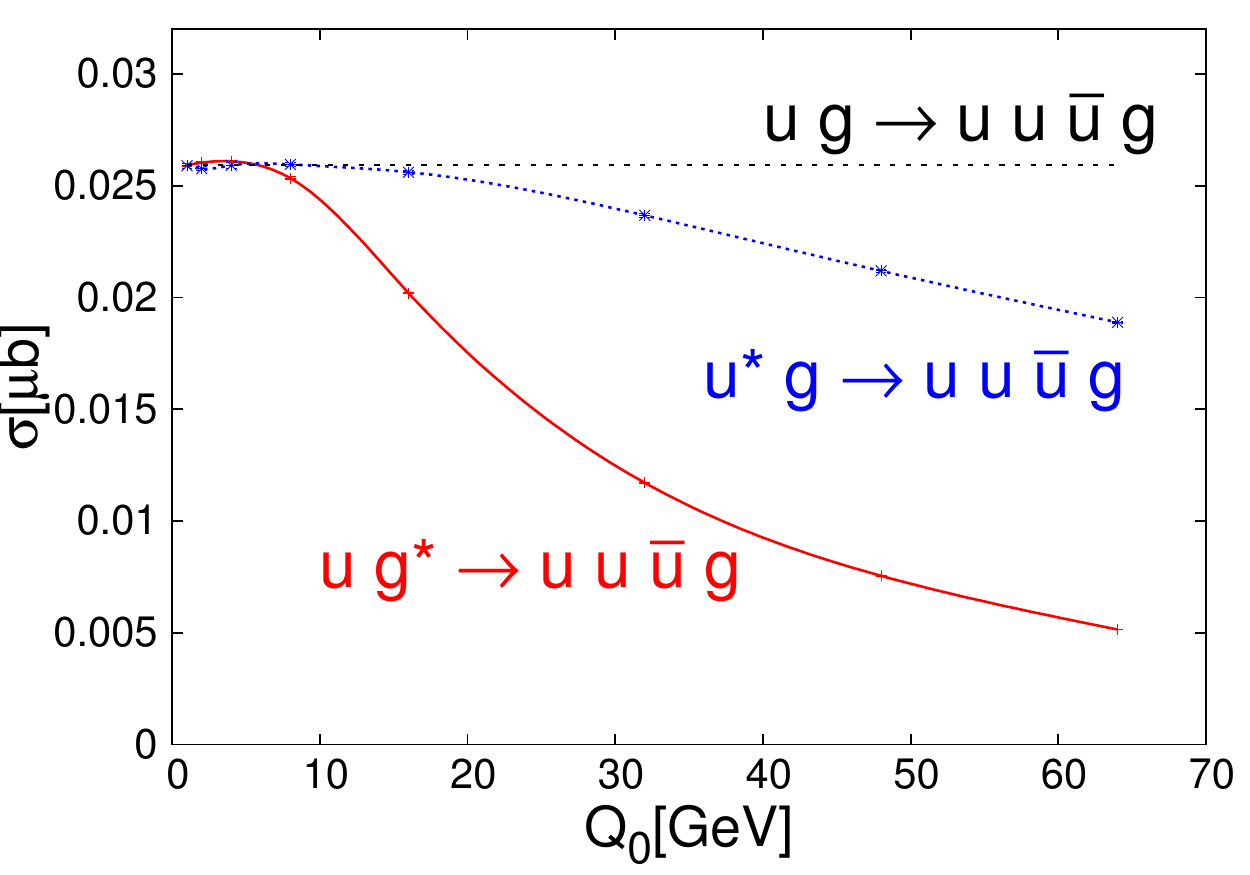,width=0.45\linewidth}
\caption{\small\label{Fig5}The cross section of various processes as function of the scale $Q_0$ in the toy model unintegrated pdf of \Equation{eqn:defUnint}. The on-shell process does not depend on this scale, and its value corresponds to $Q_0=0$. The other curves in each plot correspond to the case of an off-shell intial-state gluon and an off-shell initial-state quark.}
}
Results are presented in \Figure{Fig5}.
Depicted are cross sections as function of the scale $Q_0$.
The center-of-mass energy is $14\mathrm{TeV}$, and the phase space cuts are $p_T>20\mathrm{GeV}$ and $|y|<2.8$ for all final-state particles.
Also, all pairs of final-state particles have $\Delta R>0.4$, except in the first process.
The collinear pdfs are from CTEQ6L1~\cite{Pumplin:2002vw}.
The values of the couplings and masses are the same as in~\cite{vanHameren:2012if}, and also the scale $\mu$ is fixed to the $Z$-mass again for simplicity.

The straight lines in each of the plots correspond to the collinear case, that is to $Q_0=0$.
Since the unintegrated pdf is merely a toy model, one should be careful in drawing conclusions from the other curves.
It is, however, clear that the matrix elements for the off-shell quark behave very differently from the matrix elements for the off-shell gluon for increasing $Q_0$, that is for increasing values of the virtuality of the off-shell parton.

\section{Summary}
We presented a prescription to calculate manifestly gauge invariant tree-level scattering amplitudes for arbitrary scattering processes with off-shell initial-state quarks within the kinematics of high-energy scattering.
Furthermore, we derived explicit expressions for the helicity amplitudes of the process $u^*g\to ug$, and studied the difference in behavior of the matrix elements between processes $u^*g\to uX$ and $ug^*\to uX$ for a number of sets of final-state particles.
We see that the matrix elements for off-shell gluons suppress the cross section stronger for increasing values of the transverse momentum than the matrix elements for off-shell quarks.
%

\subsection*{Acknowledgments}
The authors would like to thank P.~Kotko for useful discussions and comments.
This work was partially supported by HOMING PLUS/2010-2/6: ``Matrix Elements and Exclusive Parton Densities for Large Hadron Collider''.

\providecommand{\href}[2]{#2}\begingroup\raggedright\endgroup

\begin{appendix}

\section{\label{appA}Reproduction of existing results}
We compare our approach with the exisiting results for the processes
%
\begin{align}
u^* \,\bar{u}^* &\to g \,g\\
u^* \,g^* &\to u \,g
~.
\end{align}
The existing results can be found  in~\cite{Nefedov:2013ywa}, where they were calculated following the effective action approach of~\cite{Lipatov:2000se}.
%
\subsection{$u^* \,\bar{u}^* \to g \,g$}
The existing result can be found in eq.~(29) and eq.~(21) of~\cite{Nefedov:2013ywa}.
\myFigure{
\epsfig{figure=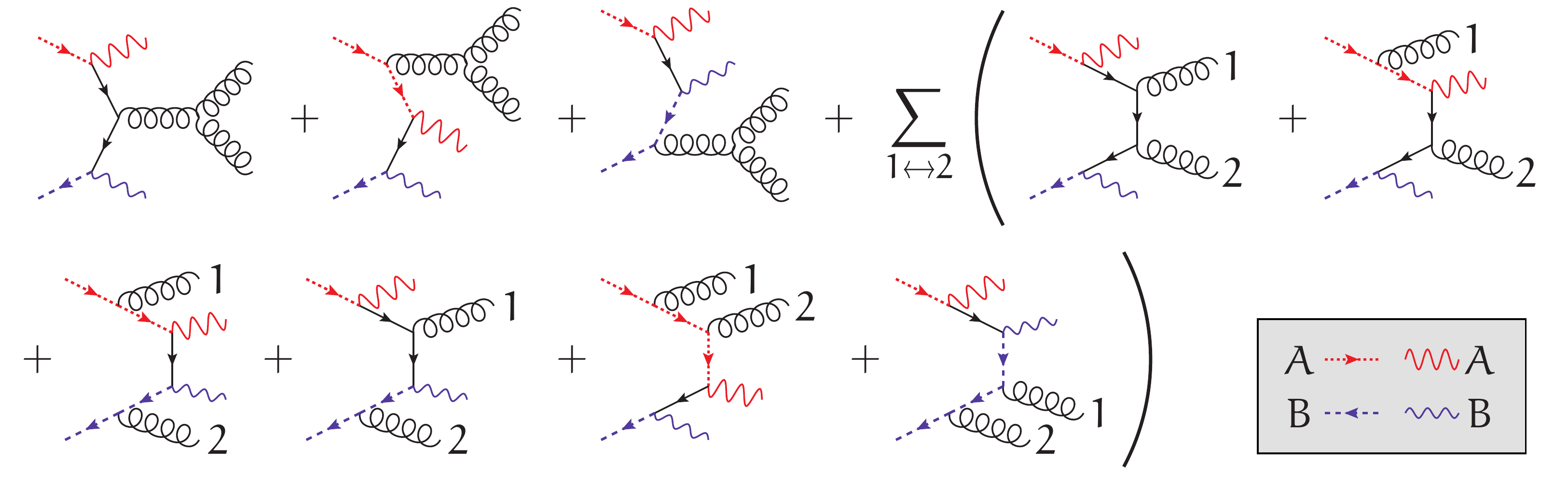,width=0.9\linewidth}
\caption{\small\label{Fig6}All graphs contributing to the process $u^*\bar{u}^*\to gg$ via the embedding $q_A\,\bar{q}_B\to\gamma_A\gamma_B\,gg$.}
}
The graphs contributing to the embedding
%
\begin{equation}
q_A(k_1)\;\bar{q}_B(k_2) \;\to\; \gamma_A\;\gamma_B\;g(p_1,a,\mu)\;g(p_2,b,\nu)
\end{equation}
%
in our approach are depicted in \Figure{Fig6}.
Using the relation
%
\begin{equation}
-\imag\gQCD^2\,T^cf^{cab} = (-\imag\gQCD)(-\imag\gQCD)[T^a,T^b]
\end{equation}
%
we split the graphs with the $3$-gluon vertex into two parts.
The amplitude can, before contraction with the polarization vectors of the external gluons, be written as
\begin{equation}
\EuScript{M}^{ab,\,\mu\nu}_{-,-} = \sqrt{\frac{x_1x_2k_1^2k_2^2}{4}}\,(-\imag\gQCD)^2\,\langle\ell_2|\Big[C^{\mu\nu}(p_1,p_2)T^aT^b+C^{\nu\mu}(p_2,p_1)T^bT^a\Big]|\ell_1]
~,
\end{equation}
and we only need to calculate $C^{\nu\mu}(p_2,p_1)$.
The spinors $|\ell_1],\langle\ell_2|$ are those assigned to the external auxiliary quarks $q_A$ and $\bar{q}_B$ respectively and both have negative helicity.
The mixed-helicity cases vanish, and the positive helicity case will be dealt with later.
According to our prescription, the polarization vectors for the auxiliary photons are then given by
%
\begin{equation}
\vep_A^\mu = \frac{\langle\ell_1|\gamma^\mu|\ell_2]}{\sqrt{2}[\ell_1|\ell_2]}
\quad,\quad
\vep_B^\mu = \frac{\langle\ell_1|\gamma^\mu|\ell_2]}{\sqrt{2}\langle\ell_1|\ell_2\rangle}
~.
\end{equation}
%
Remember that for the $B$-photon $\ell_1$ and $\ell_2$ switch role.
The $3$-gluon vertex with momentum conservation imposed and a propagator denominator included we denote by
%
\begin{equation}
V^{\sigma\mu\nu}(p_1,p_2) = \frac{1}{(p_1+p_2)^2}\big[(p_1-p_2)^\sigma\,\eta^{\mu\nu} + (2p_2+p_1)^\mu\,\eta^{\nu\sigma} - (2p_1+p_2)^\nu\,\eta^{\mu\sigma}\big]
~.
\end{equation}
%
The graphs contribute, in the Feynman gauge, to $C^{\nu\mu}(p_2,p_1)$ as follows:
%
\begin{align}
C^{\nu\mu}(p_2,p_1)
&= (-\imag\slashe_B)\frac{\imag}{-\slashk_2}(-\imag\gamma_\sigma)V^{\sigma\nu\mu}(p_2,p_1)\,\frac{\imag}{\slashk_1}(-\imag\slashe_A)
\\
&+ (-\imag\slashe_B)\frac{\imag}{-\slashk_2}(-\imag\slashe_A)\frac{\imag\slashl_1}{2\lid{\ell_1}{(k_1-p_1-p_2)}}(-\imag\gamma_\sigma)V^{\sigma\nu\mu}(p_2,p_1)
\\
&+ (-\imag\gamma_\sigma)V^{\sigma\nu\mu}(p_2,p_1)\,\frac{\imag\slashl_2}{2\lid{\ell_2}{(-k_2+p_1+p_2)}}(-\imag\slashe_B)\frac{\imag}{\slashk_1}(-\imag\slashe_A)
\\
&+ (-\imag\slashe_B)\frac{\imag}{-\slashk_2}\,\gamma^\nu\,\frac{\imag}{\slashk_1-\slashp_1}\,\gamma^\mu\,\frac{\imag}{\slashk_1}(-\imag\slashe_A)
\\
&+ (-\imag\slashe_B)\frac{\imag}{-\slashk_2}\,\gamma^\nu\,\frac{\imag}{\slashk_1-\slashp_1}\,(-\imag\slashe_A)\frac{\imag\slashl_1}{2\lid{\ell_1}{(k_1-p_1)}}\,\gamma^\mu
\\
&+ \gamma^\nu\,\frac{\imag\slashl_2}{2\lid{\ell_2}{(-k_2+p_2)}}(-\imag\slashe_B)\,\frac{\imag}{\slashk_1-\slashp_1}\,(-\imag\slashe_A)\frac{\imag\slashl_1}{2\lid{\ell_1}{(k_1-p_1)}}\,\gamma^\mu
\\
&+ \gamma^\nu\,\frac{\imag\slashl_2}{2\lid{\ell_2}{(-k_2+p_2)}}(-\imag\slashe_B)\,\frac{\imag}{\slashk_1-\slashp_1}\,\gamma^\mu\,\frac{\imag}{\slashk_1}(-\imag\slashe_A)
\\
&+ (-\imag\slashe_B)\frac{\imag}{-\slashk_2}(-\imag\slashe_A)\frac{\imag\slashl_1}{2\lid{\ell_1}{(k_1-p_1-p_2)}}\,\gamma^\nu\,\frac{\imag\slashl_1}{2\lid{\ell_1}{(k_1-p_1)}}\,\gamma^\mu
\\
&+ \gamma^\nu\,\frac{\imag\slashl_2}{2\lid{\ell_2}{(-k_2+p_2)}}\,\gamma^\mu\,\frac{\imag\slashl_2}{2\lid{\ell_2}{(-k_2+p_1+p_2)}}(-\imag\slashe_B)\frac{\imag}{\slashk_1}(-\imag\slashe_A)
\end{align}
%
After some re-arrangements of terms and using the fact that $\lid{\ell_1}{k_1}=\lid{\ell_2}{k_2} = 0$, and therefor also
%
\begin{equation}
\lid{\ell_1}{(p_1+p_2)}=\lid{\ell_1}{(k_1+k_2)}=\lid{\ell_1}{k_2}
\quad,\quad
\lid{\ell_2}{(p_1+p_2)}=\lid{\ell_2}{(k_1+k_2)}=\lid{\ell_2}{k_1}
~,
\end{equation}
we find
%
\begin{align}
-\imag &C^{\nu\mu}(p_2,p_1)
\notag\\
&= \left(
 \slashe_B\,\frac{1}{\slashk_2}\,\gamma_\sigma\,\frac{1}{\slashk_1}\,\slashe_A
-\slashe_B\,\frac{1}{\slashk_2}\,\slashe_A\,\frac{\slashl_1}{2\lid{\ell_1}{k_2}}\,\gamma_\sigma
- \gamma_\sigma\,\frac{\slashl_2}{2\lid{\ell_2}{k_1}}\,\slashe_B\,\frac{1}{\slashk_1}\,\slashe_A\right) V^{\sigma\nu\mu}(p_2,p_1)
\notag\\
&+\left( \slashe_B\,\frac{1}{\slashk_2}\,\gamma^\nu
         - \gamma^\nu\,\frac{\slashl_2}{2\lid{\ell_2}{p_2}}\,\slashe_B \right)
\frac{1}{\slashp_1-\slashk_1}
\left( \gamma^\mu\,\frac{1}{\slashk_1}\,\slashe_A
      - \slashe_A\,\frac{\slashl_1}{2\lid{\ell_1}{p_1}}\,\gamma^\mu\right)
\notag\\
&-\slashe_B\,\frac{1}{\slashk_2}\,\slashe_A\,\frac{\slashl_1}{2\lid{\ell_1}{k_2}}\,\gamma^\nu\,\frac{\slashl_1}{2\lid{\ell_1}{p_1}}\,\gamma^\mu
+\gamma^\nu\,\frac{\slashl_2}{2\lid{\ell_2}{p_2}}\,\gamma^\mu\,\frac{\slashl_2}{2\lid{\ell_2}{k_1}}\,\slashe_B\,\frac{1}{\slashk_1}\,\slashe_A
~.
\end{align}
%
It turns out to be convenient to insert $1=\slashk_1/\slashk_1$ and $1=\slashk_2/\slashk_2$ at some points:
%
\begin{align}
-\imag &C^{\nu\mu}(p_2,p_1)
\notag\\
&= \left(
 \slashe_B\,\frac{1}{\slashk_2}\,\gamma_\sigma\,\frac{1}{\slashk_1}\,\slashe_A
-\slashe_B\,\frac{1}{\slashk_2}\,\slashk_1\,\frac{1}{\slashk_1}\,\slashe_A\,\frac{\slashl_1}{2\lid{\ell_1}{k_2}}\,\gamma_\sigma
- \gamma_\sigma\,\frac{\slashl_2}{2\lid{\ell_2}{k_1}}\,\slashe_B\,\frac{1}{\slashk_2}\,\slashk_2\,\frac{1}{\slashk_1}\,\slashe_A\right) V^{\sigma\nu\mu}(p_2,p_1)
\notag\\
&+\left( \slashe_B\,\frac{1}{\slashk_2}\,\gamma^\nu
         - \gamma^\nu\,\frac{\slashl_2}{2\lid{\ell_2}{p_2}}\,\slashe_B\,\frac{1}{\slashk_2}\,\slashk_2 \right)
\frac{1}{\slashp_1-\slashk_1}
\left( \gamma^\mu\,\frac{1}{\slashk_1}\,\slashe_A
      - \slashk_1\,\frac{1}{\slashk_1}\,\slashe_A\,\frac{\slashl_1}{2\lid{\ell_1}{p_1}}\,\gamma^\mu\right)
\notag\\
&-\slashe_B\,\frac{1}{\slashk_2}\,\slashk_1\,\frac{1}{\slashk_1}\,\slashe_A\,\frac{\slashl_1}{2\lid{\ell_1}{k_2}}\,\gamma^\nu\,\frac{\slashl_1}{2\lid{\ell_1}{p_1}}\,\gamma^\mu
+\gamma^\nu\,\frac{\slashl_2}{2\lid{\ell_2}{p_2}}\,\gamma^\mu\,\frac{\slashl_2}{2\lid{\ell_2}{k_1}}\,\slashe_B\,\frac{1}{\slashk_2}\,\slashk_2\,\frac{1}{\slashk_1}\,\slashe_A
~.
\end{align}
%
Now, we apply the fact that
%
\begin{align}
\slashl_1 = |\ell_1\rangle[\ell_1| + |\ell_1]\langle\ell_1|
\quad&,\quad
\slashl_3 = |\ell_2\rangle[\ell_1| + |\ell_1]\langle\ell_2|
\notag\\
\slashl_2 = |\ell_2\rangle[\ell_2| + |\ell_2]\langle\ell_2|
\quad&,\quad
\slashl_4 = |\ell_1\rangle[\ell_2| + |\ell_2]\langle\ell_1|
\end{align}
%
and
\begin{equation}
\slashe_A = \frac{\sqrt{2}}{[\ell_1|\ell_2]}\,\slashl_4
\quad,\quad
\slashe_B = \frac{\sqrt{2}}{\langle\ell_1|\ell_2\rangle}\,\slashl_4
\end{equation}
and the general relations
%
\begin{gather}
[p|p] = \langle p|p\rangle = 0
\quad,\quad
\langle q|p] = [q|p\rangle = 0
\quad,\quad
\langle p|\gamma^\mu|q\rangle = [p|\gamma^\mu|q] = 0
\quad,
\nonumber\\
[p|q]=-[q|p]
\quad,\quad
\langle p|q\rangle=-\langle q|p\rangle
\quad,\quad
\langle p|\gamma^\mu|q] = [q|\gamma^\mu|p\rangle
\quad,
\\
\langle p|\gamma^\mu|p] = 2p^\mu
\quad,\quad
[p|q]\langle q|p\rangle = 2\lid{p}{q}
\quad.
\nonumber
\end{gather}
%
We get
%
\begin{align}
\langle\ell_2|\,\slashe_B &= -\sqrt{2}\,[\ell_2|
\\
\langle\ell_2|\,\gamma^\sigma\,\frac{\slashl_2}{2\lid{\ell_2}{k_1}}\,\slashe_B
&= - \sqrt{2}\,\frac{\ell_{2}^{\sigma}}{\lid{\ell_2}{k_1}}\,[\ell_2|
\\
\langle\ell_2|\,\gamma^\nu\,\frac{\slashl_2}{2\lid{\ell_2}{p_2}}\,\gamma^\mu\,\frac{\slashl_2}{2\lid{\ell_2}{k_1}}\,\slashe_B
&= -\sqrt{2}\,\frac{\ell_2^\nu}{\lid{\ell_2}{p_2}}\,\frac{\ell_2^\mu}{\lid{\ell_2}{k_1}}\,[\ell_2|
\end{align}
%
and likewise
%
\begin{align}
\slashe_A\,|\ell_1] &= -\sqrt{2}\,|\ell_1\rangle
\\
\slashe_A\,\frac{\slashl_1}{2\lid{\ell_1}{k_2}}\,\gamma^\sigma\,|\ell_1]
&=-\sqrt{2}\,\frac{\ell_1^\sigma}{\lid{\ell_1}{k_2}}\,|\ell_1\rangle
\\
\slashe_A\,\frac{\slashl_1}{2\lid{\ell_1}{k_2}}\,\gamma^\nu\,\frac{\slashl_1}{2\lid{\ell_1}{p_1}}\,\gamma^\mu\,|\ell_1]
&=-\sqrt{2}\,\frac{\ell_1^\nu}{\lid{\ell_1}{k_2}}\,\frac{\ell_1^\mu}{\lid{\ell_1}{p_1}}\,|\ell_1\rangle
~.
\end{align}
%
Inserting these relations, we get
%
\begin{equation}
\langle\ell_2|\,C^{\nu\mu}(p_2,p_1)\,|\ell_1]
= 2\imag\,[\ell_2|\,\frac{1}{\slashk_2}\,D^{\nu\mu}(p_2,p_1)\,\frac{1}{\slashk_1}\,|\ell_1\rangle
~,
\end{equation}
%
where
%
\begin{multline}
D^{\nu\mu}(p_2,p_1) =
\bigg(\gamma_\sigma - \slashk_1\frac{\ell_{1\sigma}}{\lid{\ell_1}{k_2}}
                    - \slashk_2\frac{\ell_{2\sigma}}{\lid{\ell_2}{k_1}}\bigg)V^{\sigma\nu\mu}(p_2,p_1)
\\
+\bigg(\gamma^\nu-\slashk_2\frac{\ell_2^\nu}{\lid{\ell_2}{p_2}}\bigg)\frac{1}{\slashp_1-\slashk_1}\bigg(\gamma^\mu-\slashk_1\frac{\ell_1^\mu}{\lid{\ell_1}{p_1}}\bigg)
-\slashk_1\frac{\ell_1^\nu}{\lid{\ell_1}{k_2}}\frac{\ell_1^\mu}{\lid{\ell_1}{p_1}}
+\slashk_2\frac{\ell_2^\nu}{\lid{\ell_2}{k_1}}\frac{\ell_2^\mu}{\lid{\ell_2}{p_2}}
~.
\end{multline}
%
One can now already recognize the terms from eq.~(21) of~\cite{Nefedov:2013ywa}, with the identifications
%
\begin{align}
\ell_1\leftrightarrow n^-
\quad,&\quad
\ell_2\leftrightarrow n^+
\\
\gamma^\mu-\slashk_1\frac{\ell_1^\mu}{\lid{\ell_1}{p_1}} \leftrightarrow \gamma^{(-)\mu}(k_1,-p_1)
\quad,&\quad
\gamma^\nu-\slashk_2\frac{\ell_2^\nu}{\lid{\ell_2}{p_2}} \leftrightarrow \gamma^{(+)\nu}(k_2,-p_2)
\\
\gamma_\sigma - \slashk_1\frac{\ell_{1\sigma}}{\lid{\ell_1}{k_2}}
                   - \slashk_2\frac{\ell_{2\sigma}}{\lid{\ell_2}{k_1}}
&\leftrightarrow
\gamma^{(+,-)}_\sigma(k_1,k_2)
\quad.
\end{align}
%
More precisely, and remembering that $p_1-k_1=k_2-p_2$, we find
%
\begin{equation}
C^{gg,\,ab,\,\mu\nu}_{Q\bar{Q}}(k_1,k_2,p_1,p_2)
=\gQCD^2\,\big[ D^{\mu\nu}(p_1,p_2)T^aT^b + D^{\nu\mu}(p_2,p_1)T^bT^a \big]
~.
\end{equation}
%
So we arrived at
%
\begin{equation}
\EuScript{M}^{ab,\,\mu\nu}_{-,-} = -\imag\sqrt{x_1x_2k_1^2k_2^2}\,[\ell_2|\,\frac{1}{\slashk_2}\,C^{gg,\,ab,\,\mu\nu}_{Q\bar{Q}}\,\frac{1}{\slashk_1}\,|\ell_1\rangle
~.
\end{equation}
%
Writing $k_1$ and $k_2$ in terms of $\ell_{1,2,3,4}$ and using $[\ell_2|\slashl_2=[\ell_2|\slashl_4 = \slashl_1|\ell_1\rangle=\slashl_4|\ell_1\rangle=0$, we have
%
\begin{align}
[\ell_2|\,\frac{1}{\slashk_2}
&=\frac{1}{k_2^2}\,[\ell_2|\,\slashk_2
=\frac{1}{k_2^2}\,[\ell_2|\,\slashl_3\,\frac{-\lid{k_{2\perp}}{\ell_4}}{\lid{\ell_1}{\ell_2}}
=\frac{1}{k_2^2}\,[\ell_2|\ell_1]\langle\ell_2|\,\frac{-\lid{k_{2\perp}}{\ell_4}}{\lid{\ell_1}{\ell_2}}
\\
\frac{1}{\slashk_1}\,|\ell_1\rangle
&=\frac{1}{k_1^2}\,\slashk_1\,|\ell_1\rangle
=\frac{1}{k_1^2}\,\frac{-\lid{k_{1\perp}}{\ell_4}}{\lid{\ell_1}{\ell_2}}\,\slashl_3\,|\ell_1\rangle
=\frac{1}{k_1^2}\,\frac{-\lid{k_{1\perp}}{\ell_4}}{\lid{\ell_1}{\ell_2}}\,|\ell_1]\langle\ell_2|\ell_1\rangle
\label{eq:1116}
\end{align}
%
Using furthermore that
$[\ell_2|\ell_1]\langle\ell_2|\ell_1\rangle=-2\lid{\ell_1}{\ell_2}$ and $\sqrt{x_2}\langle\ell_2|=\langle x_2\ell_2|$ and $\sqrt{x_1}|\ell_1]=|x_1\ell_1]$
we find
%
\begin{equation}
\EuScript{M}^{ab,\,\mu\nu}_{-,-} = \imag\sqrt{\frac{1}{k_1^2k_2^2}}\,
\frac{2\,\lid{k_{1\perp}}{\ell_4}\,\lid{k_{2\perp}}{\ell_4}}{\lid{\ell_1}{\ell_2}}\,\langle x_2\ell_2|\,C^{gg,\,ab,\,\mu\nu}_{Q\bar{Q}}\,|x_1\ell_1]
~.
\end{equation}
%
And thus we established the equality with eq.~(29) of~\cite{Nefedov:2013ywa} up to a phase factor, as is clear from \Equation{eqn:kperpsquare}.
The reader may convince themselves that
%
\begin{equation}
\EuScript{M}^{ab,\,\mu\nu}_{+,+} = \imag\sqrt{\frac{1}{k_1^2k_2^2}}\,
\frac{2\,\lid{k_{1\perp}}{\ell_3}\,\lid{k_{2\perp}}{\ell_3}}{\lid{\ell_1}{\ell_2}}\,[x_2\ell_2|\,C^{gg,\,ab,\,\mu\nu}_{Q\bar{Q}}\,|x_1\ell_1\rangle
~.
\end{equation}

\subsection{$u^* \,g^* \to u \,g$}
The existing result can be found in eq.~(24) and eq.~(16) of~\cite{Nefedov:2013ywa}.
The graphs contributing to the embedding
\begin{equation}
q_A(k_1)\,q_B(k_2)\;\to\; \gamma_A\,q_B\,u(p_1)\,g(p_2,\mu,b)
\nonumber
\end{equation}
%
are depicted in \Figure{Fig7}.
\myFigure{
\epsfig{figure=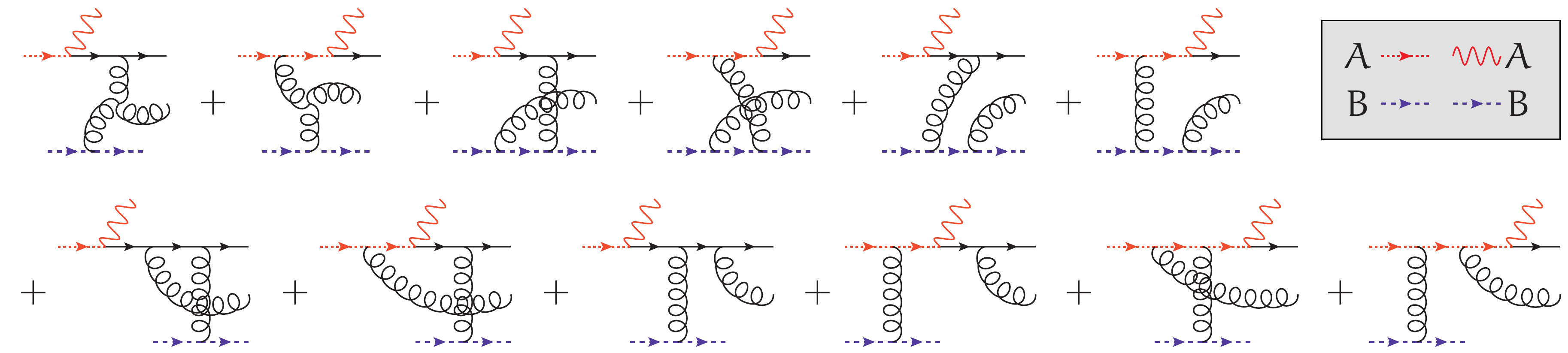,width=0.99\linewidth}
\caption{\small\label{Fig7}All graphs contributing to the process $u^*g^*\to ug$ via the embedding $q_A\,q_B\to\gamma_A\,q_B\,u\,g$.}
}
Rather than going through the whole calculation in all detail, we identify which graphs contribute to which terms in eq.~(16) of~\cite{Nefedov:2013ywa}.

First of all, we recognize that all graphs except the last two can be paired according to the occurrence of the combination
%
\begin{align}
&\hspace{-1ex}\epsfig{figure=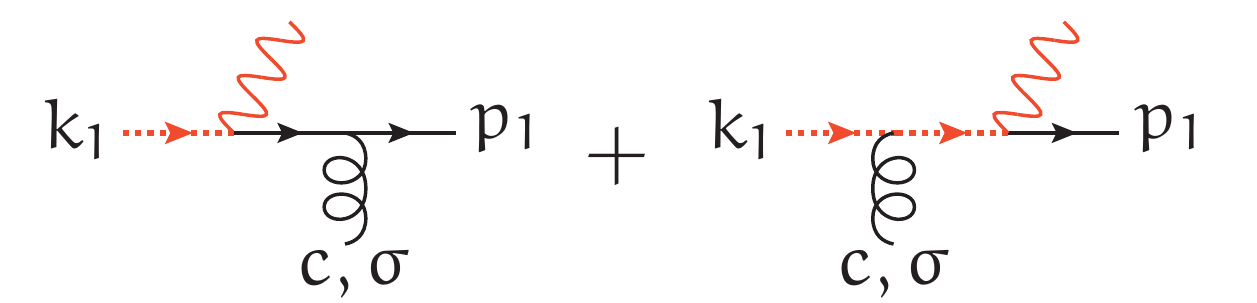,width=36ex}
\notag\\
&=\;
(-\imag\gamma^\sigma\,T^c)\frac{\imag}{\slashk_1}(-\imag\slashe_A)|\ell_1]
\;+\;
(-\imag\slashe_A)\frac{\imag\slashl_1}{2\lid{\ell_1}{p_1}}(-\imag\gamma^\sigma\,T^c)|\ell_1]
\notag\\
&=\;
\sqrt{2}\,\imag\left(
\gamma^\sigma\frac{1}{\slashk_1}
+
\frac{\ell_1^\sigma}{\lid{\ell_1}{p_1}}
\right)|\ell_1\rangle\,T^c
=
\sqrt{2}\,\imag\,\gamma^{(-)\sigma}(k_1,p_1)\,\frac{1}{\slashk_1}|\ell_1\rangle\,T^c
\label{eq:1186}
\end{align}
with the {\em induced vertex}
%
\begin{equation}
\gamma^{(-)\sigma}(k_1,p_1)
=\gamma^\sigma
+
\slashk_1\,\frac{\ell_1^\sigma}{\lid{\ell_1}{p_1}}
=\gamma^\sigma
+
\slashk_1\,\frac{\ell_1^\sigma}{\lid{\ell_1}{(p_1-k_1)}}
~.
\end{equation}
%
The factor $\slashk_1^{-1}|\ell_1\rangle$ evaluates further following \Equation{eq:1116}.
Furthermore, we recognize in the last four graphs of the first line of \Figure{Fig7} another {\em induced vertex} 
%
\begin{align}
&\hspace{-1ex}\epsfig{figure=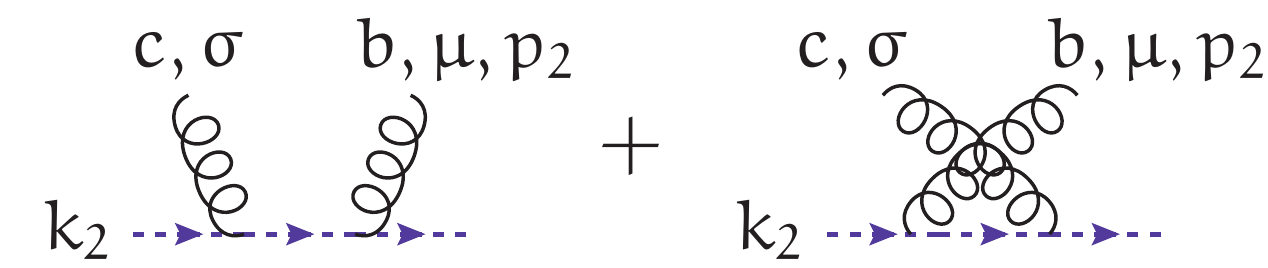,width=36ex}
\notag\\
&=\;
\langle\ell_2|(-\imag\gamma^\mu T^b)\frac{\imag\slashl_2}{2\lid{\ell_2}{p_2}}(-\imag\gamma^\sigma\,T^c)|\ell_2]
\;+\;
\langle\ell_2|(-\imag\gamma^\sigma\,T^c)\frac{\imag\slashl_2}{2\lid{\ell_2}{(k_2-p_2)}}(-\imag\gamma^\mu T^b)|\ell_2]
\\
&=\;
-2\imag\,\frac{\ell_2^\mu\ell_2^\sigma}{\lid{\ell_2}{p_2}}\,[T^b,T^c]
=
2\,\frac{\ell_2^\mu\ell_2^\sigma}{\lid{\ell_2}{p_2}}\,f^{abc}T^a
~.\notag
\end{align}
%
So all graphs on the first line have the same color factor, and we may remove the $T^a$ coming from our rules~\cite{vanHameren:2013gba} from all graphs.
The color index $a$ then indicates the off-shell initial-state gluon.
It is then straightforward to see that the whole first line of \Figure{Fig7} contributes the terms
%
\begin{equation}
\frac{1}{k_2^2}\,\gamma^{(-)\sigma}(k_1,p_1)\frac{1}{(k_1-p_1)^2}\left(\gamma^{\mu\nu\sigma}(p_2,-k_2)\ell_{2\nu}
+k_2^2\,\frac{\ell_2^\mu\ell_2^\sigma}{\lid{\ell_2}{p_2}}\right)[T^a,T^b]
\end{equation}
%
from the expression of~\cite{Nefedov:2013ywa}.
The extra factor $\sqrt{k_2^2}$ that has to be provided according to our rules reduces the factor $k_2^2$ in the denominator to $\sqrt{k_2^2}$.
Realize that $\ell_{1,2}$ are dimensionful contrary to $n^\pm$.
This difference manifests itself in the overall factor $q_2^{-}$ in the expression of~\cite{Nefedov:2013ywa} which does not occur here.
This factor also contains the $x_2$ that has to be provided separately according to our rules.

One can also easily recognize that the first two graphs on the second line of \Figure{Fig7} contribute
\begin{equation}
-\frac{1}{k_2^2}\,\slashl_2\,\frac{1}{\slashk_1-\slashp_2}\,\gamma^{(-)\mu}(k_1,-p_2)T^aT^b
~,
\end{equation}
%
and the third and the fourth graph contribute
%
\begin{equation}
-\frac{1}{k_2^2}\,\gamma^\mu\,\frac{1}{\slashk_1+\slashk_2}\,\gamma^{(-)\sigma}(k_1,k_2)\ell_{2\sigma}T^bT^a
~.
\end{equation}
%
The last two graphs, finally, can easily be seen to contribute
%
\begin{equation}
\frac{1}{k_2^2}\,\frac{2\slashk_1\ell_1^\mu}{\lid{\ell_1}{p_1}}
\left(\frac{T^aT^b}{\lid{\ell_1}{p_2}}-\frac{T^bT^a}{\lid{\ell_1}{k_2}}\right)
~.
\end{equation}
%
The factor $\slashk_1$ facilitates a compensating $\slashk_1^{-1}$ necessary to write the contribution of the graphs such that they contain $\slashk_1^{-1}|\ell_1\rangle$ following \Equation{eq:1186}.

\end{appendix}

\end{document}